\documentclass[a4paper,10pt,oneside]{article}

\usepackage{amsmath}
\usepackage{amsthm}
\usepackage{amssymb}
\usepackage{bbold}
\usepackage{booktabs}
\usepackage{cite}

\usepackage{braket} 

\usepackage{type1ec}
\usepackage[T1]{fontenc}
\usepackage[utf8]{inputenc}
\usepackage{layaureo}
\usepackage[english]{babel}

\usepackage{csquotes}
\usepackage{titling}

\title{Duality and measurement: the Copenhagen reconciliation}
\author{Vincenzo Chilla}
\date{}

\begin{document}

\maketitle

\begin{abstract}

Duality, not monism, constitutes the hermeneutic lens that characterizes the original Copenhagen interpretation of Quantum Mechanics. Therefore, evoking the principles of correspondence and complementarity, in this work we re-assert a dual-aspect reading of quantum theory, structured through a multi-perspective schema encompassing its ontological, analytical, epistemological, causal, and information dimensions. We then show how this schema dissolves the so-called ``measurement problem'', along with the associated ``knowledge-information'' and ``macro-micro'' dichotomies, issues historically raised within later monistic or universalist philosophical settings that ultimately depart from the traditional Copenhagen spirit.



\end{abstract}

\section{Introduction}
\label{intro}

A century after the birth of Quantum Mechanics, the academic term \emph{Copenhagen interpretation} still lacks a clear consensus, partly because its meaning has evolved throughout history~\cite{copenhagen1,copenhagen2,copenhagen3}. However, it was originally introduced by W. Heisenberg in the 1950s~\cite{heisenberg} with the intention of encompassing a coherent body of interpretative ideas related to Atomic Physics. These ideas stemmed from the work of N. Bohr and the research activity of the Institute for Theoretical Physics at the University of Copenhagen (the ``Copenhagen School''), which Bohr directed from 1921~\cite{dirb}. The key figures of this interpretation include the aforementioned Bohr and Heisenberg, as well as M. Born, P. Dirac,  P. Jordan, W. Pauli, L. Rosenfeld, and others~\cite{copmembri}. Thus, despite this wide diversity of perspectives, intellectual approaches, and personal backgrounds -- frequently at odds with each other, if not in outright opposition -- it is possible to outline the minimal set of common, foundational elements that constitute the mother of all ``orthodox'' interpretations of Quantum Mechanics~\cite{core0,core1,core2,core3}:
\begin{enumerate} 
\item emphasis on the \emph{role of the observer} in the quantum realm;
\item importance of \emph{language} and \emph{analytical dimension} of quantum description;
\item \emph{completeness} of the theory and \emph{intrinsic necessity of the classical model};
\item \emph{objective indeterminacy} and \emph{probabilistic character} of measurement; 
\item \emph{complementarity} and \emph{phenomenological contextuality}.
\end{enumerate}
Although they may occasionally be vague, these elements constitute a crucial baseline for any interpretation to be regarded as Copenhagen-inspired or as its descendant (sometimes named ``Copenhagen-like interpretations'' in the literature).

The mathematical formalization of Quantum Mechanics, which has made possible a remarkable predictive and applicative power of the theory, is due to the Hungarian mathematician J. von Neumann~\cite{vneumann2}. This formalism is still taught today in every basic university course on the topic, and it provided the essential framework for all other mathematical developments within the theory. In the philosophical assumptions of von Neumann's treatment, all the constitutive elements of the ``Copenhagen spirit'' can be discerned, albeit with some unresolved internal tension -- notably between a universalist conception of the physical world, required for a general and consistent quantum description, and the necessary \emph{classicality} of the apparatus of amplification and recording, as prescribed by the Copenhagen doctrine
~\cite{univ}. Operationally, this tension translates into the conceptual distinction between the designed \emph{apparatus} and the \emph{measured object}, via the \emph{Heisenberg cut}, which in von Neumann's formalism becomes a crucial distinction between the \emph{observer} and the \emph{observed universe}, via the \emph{von Neumann cut}. Although these ``cuts'' are treated as synonyms in the literature, distinguishing between them is significant to our context: for the Copenhagen School the cut has, belonging to the apparatus, an inherent \emph{physical} connotation; whereas for von Neumann it has, belonging to the observer, an \emph{analytical} (specifically \emph{logical}) one. Consequently, the assumed arbitrariness in the choice of its position also acquires a different meaning: for the Copenhagen School, it is expressive of the experimenter's freedom of choice within the measurement setup; whereas in von Neumann's view, it constitutes a logical conventionality necessary for a coherent physical description. Clearly, von Neumann is aware of this tension. In fact, formulating the problem in the notorious chapter VI of his treatise (devoted to measurement), by considering the observer as a human, he evokes the ``principle of psycho-physical parallelism'' as ``a fundamental requirement of the scientific viewpoint'' and as a possible solution for the tension. Indeed, by this principle, he can state that~\cite{vneumann2} 
\begin{displayquote}
it must be possible so to describe the extra-physical process of subjective perception as if it were in the reality of the physical world -- \emph{i.e.}, to assign to its parts equivalent physical processes in the objective environment, in ordinary space. 
\end{displayquote}
The problem is that, from the point of view of the Copenhagen interpretation, the ``principle of psycho-physical parallelism'' cannot provide such a solution, since in principle the observer does not necessarily have a ``psychic nature''\footnote{Von Neumann's powerful formalism has been indeed embedded over time in the Copenhagen interpretation, but the significance of its philosophical assumptions, especially in relation to the physical content, perhaps has not been investigated in depth. This attitude is often summarized by the expression: ``shut up and calculate!'', which perhaps misrepresents the original philosophical spirit of the School.}~\cite{nopsy}. 

In this work, we replace such a ``principle of psycho-physical parallelism'' with a \emph{principle of physical-analytical synergy}, and we show how it more faithfully adheres to the Copenhagen vision. We can state the principle as follows: 
\begin{displayquote}
\emph{Despite the mutual irreducibility between the physical observed universe and the analytical (i.e., relational and communicative) observer, there is a full synergy linking them that, by means of the measuring process, enables the flow of physical-analytical information and yields knowledge. }
\end{displayquote}
In fact, Quantum Measurement Theory aims to describe precisely the characteristics of this flow~\cite{qmth}. This formulation can capture more effectively the physical-analytical tension highlighted above and translates it into a principle of \emph{duality} -- plausibly, Bohr would have preferred the term ``complementarity'' -- that is in no way understood as ``dualism'', that is, as an irreconcilable opposition. Indeed, it is possible to explore the meaning and implications of such a duality, and explain it in a multi-perspective way to obtain general guidelines for compliance with the original Copenhagen approach. For this purpose, the argument is organized as follows.  In section \ref{dual}, the most philosophical one, we read the five foundational interpretative elements, given at the beginning of the present introduction, in light of the principle of physical-analytical synergy. Some reflections on the formalism introduced by von Neumann are then elaborated. In section \ref{measure}, we delineate a mathematical treatment of measurement that respects the requirement of classicality of the amplification and recording apparatus. The core of the treatment consists of the Hilbert-space representation of the classical states of the apparatus, suitable for describing the interaction with the measured quantum object. In section \ref{paradoxes}, we address, according to the Copenhagen School, two celebrated (and notorious) quantum paradoxes on measurement, the ``Schr\"odinger's cat'' paradox and the ``Wigner's friend'' paradox; every interpretative proposal of Quantum Mechanics cannot avoid addressing them. Finally, in section \ref{concl}, we summarize the work and present some final considerations and remarks. 

\section{The Copenhagen criterion of duality}
\label{dual}

Famously, when speaking of the origins of the Copenhagen interpretation, Heisenberg emphasized its paradoxical connotation~\cite{copeparadox}: 
\begin{displayquote}
The Copenhagen interpretation of quantum theory starts from a paradox. Any experiment in physics, whether it refers to the phenomena of daily life or to atomic events, is to be described in the terms of classical physics. The concepts of classical physics form the language by which we describe the arrangement of our experiments and state the results. We cannot and should not replace these concepts by any others.
\end{displayquote}
With these words, an indissoluble bond is established between the physics of the experiment and the language with which the observer performs and describes it. This is, in other words, the bond between the physical existence of the observed universe and the analytical existence of the observer, two truly distinct existences which are not reducible to each other. Obviously, here, by ``classical physics'' Heisenberg means the physical theory expressed in the language derived from the human \emph{mindset}, but humanity is not a necessary condition for observation: \emph{any analytical (relational and communicative) subsistence performing an observation process on a measured object is an observer for it}. Indeed, \emph{the nature of the observer is precisely the analytical activity of language which enables the physical act of measurement}. This is the starting point of the principle of physical-analytical synergy stated above, but for the understanding of it as a duality, we need a deeper insight into the ontological distinction between observed and observer. 

\subsection{Ontological duality: observed or observer}
\label{onto}

Observation, as an analytical activity, is a logical-communicative act requiring the involvement of \emph{multiple} co-essential subsistences. These are primarily the \emph{observer} and the \emph{measurer}, who together form an analytical community. The activity of observation is distinct from measurement itself: the observer does not perform the measurement but is simply \emph{informed} of its outcome, produced by the \emph{measurer}. Thus, \emph{the measurer is observed} in the act of performing the measurement. In parallel, a measurer cannot exist without an observer for whom the measurement is performed. The activity of measurement is therefore a fundamentally logical act, dependent on a shared \emph{logos} within this analytical community. This logos can be termed \emph{measurement logos}, as it establishes a logical \emph{inquiring unity} that \emph{incorporates} the measurer into the observer\footnote{As a consequence of this unity, for a given experimental context, there can be only one active measurer. Any other analytical component within the same unity must therefore act as an observer.}.  Although this may seem abstract, it is central to the Copenhagen interpretation. When applied to the human analytical community, it manifests as Bohr's concern for the logical communicability of the measurement procedure and its outcome, which in turn implies a social and ethical value~\cite{bohrlogos}.

In the measurement process, the analytical subsistence of the measurer assumes a physical nature, manifesting as the \emph{measurement apparatus}. This establishes an ontological relationship between two distinct physical subsistences: the formed apparatus and the \emph{object of inquiry}. Consequently, the apparatus is observed but not measured, while the object of inquiry is both observed and measured. The crux of the process lies in the measurer's \emph{duality} of natures: a primordial analytical nature, which it shares with the observer, and a physical nature assumed for the measurement logos, which it shares with the object. It is precisely this \emph{physical-analytical synergy} that enables the analytical subsistence of the measurer to \emph{embody} the physical apparatus, thus actualizing its physical activity\footnote{In the case of a human measurer, this parallels the typical model of human person, who is a single subsistence possessing a dual aspect: an analytical nature (the ``mind'') and a physical nature (the ``body'').}.

The dual ontological schema outlined above allows for a better understanding of the \emph{cut}, presenting two parallel definitions. For Heisenberg, the cut conceptually indicates the \emph{physical} boundary of demarcation between the apparatus -- embodied by the measurer -- and the object of inquiry; instead, for von Neumann, the \emph{analytical} boundary of demarcation between the measurer -- incorporated in the observer -- and the \emph{measured}, as two distinct analytical subsistences. Thus, as a reflection of the unity of the observer and the measurer in the same measurement logos, \emph{the apparatus possesses a tropos-existential actuality in relation to that same logos\footnote{In this context, we prefer to remain faithful to the philosophical term \emph{tropos} in reference to the effective mode of existence.}}. In other words, the apparatus is \emph{in act} during the measurement process. The \emph{existential tropos} of the object of inquiry, by contrast, is \emph{potential} until the measurement is complete. The \emph{completion}\footnote{Here and in what follows, the term \emph{completion} translates the Greek \emph{tele\'{\i}\=osis}, signifying not merely the conclusion of the interaction process between the apparatus and the object of inquiry, but rather the attainment of the \emph{logical} actuality of the measurement.} of the measurement occurs only when the physical interaction between apparatus and object is finished, \emph{and} the primordial analytical relationship between measurer and measured is thereby \emph{realized within the object}. This actualization of the object tropos is what constitutes a \emph{physical phenomenon}\footnote{The experimental conditions establish the existential tropos of the object of inquiry. For example, in the case of the double-slit experiment, depending on such conditions, the photon manifests on the screen in a ``wave mode'' (as interference fringes) or in a ``particle mode'' (as a single spot).}. In the specific case of humans, the necessary logical unity between observer and measurer, for the purpose of measurement, constitutes the foundation of the \emph{classicality} of the apparatus in conformity with the inquisitive mindset. This is an indispensable requirement that Bohr elevates to the \emph{correspondence principle}~\cite{corrisp}.

\subsection{Analytical duality: formal or context-natural}

In this subsection, we pay attention to the analytical dimension of the Copenhagen interpretation, focusing on the characteristics of \emph{human} language in the description of physical phenomena. Indeed, a significant concern for the Copenhagen School is addressing the ambiguities inherent in this descriptive language, seeking to resolve or dissolve them~\cite{ambiguity}. At this analytical level as well, the argument can be presented from a dual perspective by analyzing the conceptual distinction between \emph{formal language} and \emph{context-natural language}.

Formal language is a human language typically structured according to a system of mathematical logic. It  represents a more refined form of language than common language, and it is used to describe observed or measured reality (i.e., physical reality) and to communicate its characteristics. Indeed, any empirical event can be expressed as a logical proposition, and any comprehensive theory of physical phenomena must therefore be constructed using such a formal language. The archetypal formal language for the human classical mindset is \emph{Boolean logic}, which is characterized by several constitutive principles: the bivalence of truth values (true or false); the principle of the excluded middle; the universal compatibility of propositions with respect to AND and OR operations; and the property of distributivity. In contrast, \emph{quantum logic}~\cite{birkhoff} provides another example of a formal language, one that generalizes Boolean logic in significant ways. It is not strictly bivalent; the principle of the excluded middle does not universally apply; and neither the universal compatibility of propositions nor the principle of distributivity holds in general\footnote{Technically, classical logic is based on a distributive Boolean lattice, whereas quantum logic is based on a non-distributive orthomodular lattice which generalizes the former~\cite{birkhoff}.}. These differences are the key reason for the analytical ambiguities found within classical language when it is used for the description of the atomic world.

Following from the application of the principle of physical-analytical synergy and from the ontological duality as outlined above, it is expected that the formal language itself can be \emph{physically} connoted when used to describe the experimental process. It is in this way that it corresponds to a context-natural language, in which the physical properties of the object, emerging from the measurement, are indeed expressed by logically true propositions. In other words, in the context-natural language, the truth value of the proposition expressing the physical property of the object of inquiry is determined \emph{by rigorously explicating, via classical logic, the experimental context defined by the measuring apparatus and the measurement logos}. \emph{In the absence of such an explication, the property remains merely formal, as presumed by the classical mindset, and cannot be attributed to the object}. In the Copenhagen view, \emph{attributing a merely formal property -- one without context-natural meaning -- to the object of inquiry is the primary source of analytical ambiguity}~\cite{ambiguity}. Conversely, non-physical attributes, which are sometimes required for a full and adequate description of the phenomenon, can only be articulated in the formal language. A prototypical example of a non-physical, and therefore merely formal, attribute is ``non-existence''. It cannot in any way be assigned to the object of inquiry, since no experimental context can be implemented to manifest it. However, it can be used in describing the functioning of a classical apparatus (e.g., to indicate the ``absence of clicks'' in a detector).

Once the apparatus and the object of inquiry, along with their respective roles, have been clearly defined within the experimental process, and consequently, the context and the phenomenon have been delineated, it is no longer possible to make contextual changes without, in general, introducing descriptive ambiguities. As a significant case, to ask what the ``quantum properties of the instrument in a \emph{given} experiment'' are would lead to a profound and intractable analytical ambiguity, as it conflates the apparatus with the object of inquiry. Indeed, the quantum properties of the ``instrument-object'' can be properly contextualized only by defining a \emph{new} experiment  that highlights them. This would therefore be a different phenomenon from the original one. The ambiguity lies in the ill-defined position of the cut, which stems from the confusion between the observer and the measurer. Rather than being placed between the measurer and the measured (thereby distinguishing between the apparatus and the object), the cut is mistakenly positioned between the observer and the measurer. This separation breaks the inquiring unity and misconstrues the measurer as measured instead of simply observed, ultimately conflating the apparatus with the object. 

On the other hand, if the apparatus is not confused with the object of inquiry, it can still be modified, for example, by adding or removing parts. In this case, a change is implemented in the experimental context that generally alters the object's existential tropos -- its very mode of existence. This peculiar effect can be interpreted as a \emph{physical-analytical} ambiguity, also known as \emph{indeterminacy}, corresponding to an indefinite physical property~\cite{indeter}. Such an ambiguity is resolved only by measurement -- that is, by actualizing the object potential existence as observed, in relation to the measurement logos expressed in the apparatus. Therefore, for the Copenhagen interpretation, \emph{the act of measurement is the preeminent means of physical-analytical disambiguation}.

\subsection{Epistemological duality: classical or quantum}

Given a contextual phenomenon, the cut also establishes an \emph{epistemological} distinction between the description of the apparatus and the object of inquiry: the former is subject to the \emph{classical} model, while the latter (the \emph{system}) is subject to the \emph{quantum} model. It is therefore necessary to identify the peculiarities of the two descriptions in order to unify the conceptual and mathematical framework. The starting points are the concepts of the \emph{observable} and the \emph{state}, which serve to formalize specific ontological aspects of the \emph{observed} reality. These are the so-called ``elements of reality'', more precisely \emph{physical} reality, which represent the ontological building blocks of the theory.

In the classical model, the notion of an observable corresponds to the ontological idea of a real, objective \emph{physical manifestation}, and its measurement is associated with a \emph{real} number representing the instrument reading. In the specific context of mechanics, the position $q$ and the momentum $p$ of a particle constitute paradigmatic examples of observables, and the ordered pair $(q, p)$ is called the \emph{state}, i.e., an element in \emph{phase space}. The notion of state is a formal aid with no particular ontological meaning; this meaning resides solely in the observables $q$ and $p$, from which the state is formally constructed. In this framework, any observable can be defined as a real-valued function on the phase space. However, this scheme is no longer applicable in the quantum model due to the \emph{uncertainty principle}: the impossibility of simultaneously measuring the observables $q$ and $p$ renders the definition of a state as the pair $(q,p)$ meaningless\footnote{This point constitutes a significant and plausible misunderstanding in the EPR paper~\cite{epr}. The wave function of the free particle with definite momentum could be considered as the formal description of an observable in the classical fashion, as it is a function (albeit non-real) of $q$ and $p$. In the quantum model, however, it is instead the formalization of the state. \label{epr}}. To overcome this difficulty, the quantum model introduces the notion of a \emph{state space} (e.g., a Hilbert space), defining the concept of a state \emph{independently} of any specific observable. This choice strongly suggests that, contrary to the typical classical treatment outlined above, \emph{the formal notion of the quantum state actually corresponds to an ontological notion, distinct and independent from the physical manifestation associated with the observable}~\cite{ontics}. In other words, within the Copenhagen framework, the ontological model implied by quantum formalization is richer than the one implied by the classical model, as it includes new ``elements of reality'' (the states) in addition to the physical manifestations (the observables). Such a specific ontological element associated with the state can here be named \emph{character}\footnote{Thus, a third type of real distinction, in addition to those already highlighted -- observed-observer and potentiality-act -- is implicated in physical reality; it can be philosophically grounded in the traditional distinction between substance and existence~\cite{philosophy}.}.

To capture and formalize the experimental evidence of \emph{superposition} -- the true distinctive feature of the quantum model compared to the classical one -- the theory assumes that the state space is a \emph{complete complex Hilbert space}. This allows for the definition of an observable, in a manner analogous to the classical case, as a suitable \emph{Hermitian operator acting on the states}. The \emph{analytical} process of identifying a character through its physical manifestation upon measurement is captured, within the theory formalism, by the so-called \emph{eigenstate-eigenvalue link}~\cite{eelink}. This link dictates that the ideal measurement of an observable on a system in an eigenstate will yield the corresponding eigenvalue with certainty\footnote{For simplicity, the ideal case of projective measurement is here assumed. The Hermiticity of the operator associated with an observable guarantees that its eigenvalues are real numbers.}. Thus, in accordance with the principle of physical-analytical synergy, the state of the system, as an eigenstate, recapitulates both its \emph{ontic} physical manifestation upon measurement and the effect of the \emph{epistemic} analytical activity of the measurer, which attributes the corresponding eigenvalue. In fact, it is possible to associate with each eigenvector $\ket{x}$, with eigenvalue $x$, the \emph{projector} operator $\mathbb{P}_x = \ket{x} \bra{x}$, which corresponds to the \emph{experimental proposition} $\mathfrak{p}_x$: ``the system is in the state $\ket{x}$ with measured value $x$''. It is significant to note that while the projector formalizes a logical-operational description of the state, it should not be confused with the concept of an observable: a projector cannot be measured, and its eigenvalues cannot be the context-natural result of a measurement\footnote{However, interestingly, von Neumann's framework occasionally leads to confusion between states and observables. For example, in~\cite{birkhoff}, after the enunciation of the postulate for the logical conjunction of experimental propositions, a significant remark states: ``This postulate would clearly be implied by the not unnatural conjecture that all Hermitian-symmetric operators in Hilbert space (phase-space) correspond to observables''. This possible confusion follows the one between observable and state, inherited from the classical framework, already indicated in note \ref{epr}.}. Therefore, the logical compatibility of two experimental propositions is reduced to the commutativity of their respective projectors. A \emph{complete set of compatible observables}, which admits a common orthogonal basis of eigenvectors for the state space, then provides the classical Boolean core of compatible experimental propositions, around which it is possible to construct quantum logic that accounts for the superpositions of states~\cite{birkhoff}.

On the other hand, the need to involve the observed apparatus in the description of measurement, so that the principle of physical-analytical synergy also applies to it, entails that the eigenstate-eigenvalue link is applicable to the classical model as well. This means that \emph{the classical model also needs to be based on a Hilbertian formalism to describe the states of the apparatus}, even if it does not admit superposition and is thus unable to distinguish the ontology of the state from that of the observable. How can one give physical meaning to a classical Hilbertian state space while excluding linear superpositions? To answer this question, let us suppose that the apparatus has $n$ possible readings $x_i$ ($i=1, \ldots, n$) corresponding to the measured observable, and that each of these readings corresponds to a labeled \emph{stationary} trajectory in its phase space. This set of trajectories can be put into a one-to-one correspondence with a specific and \emph{fixed orthonormal} basis $\{\ket{x_i}\}$ (the ``pointer states'') of a suitable complex Hilbert space, which defines the classical Hilbert space $\mathcal{H}_{app}$ for the states of the apparatus\footnote{The so-called initial ``ready state'' of the apparatus cannot be included among the pointer basis vectors, as it does not correspond, in this context, to any stationary trajectory in its phase space. The stochastic and non-stationary regime in the operation of the apparatus is considered negligible here.}. Furthermore, the orthogonality of the basis vectors ensures an unequivocal distinction between the classical stationary trajectories. Therefore, similarly to the quantum case, the experimental proposition that the apparatus is in a state belonging to the subset $S \subseteq \{ 1, \dots, n \}$ of possible states is represented by the classical projector $\mathbb{P}_S = \sum_{i \in S} \ket{x_i} \bra{x_i}$. In this way, any observable $f$ of the apparatus is represented by a real function defined on the stationary trajectories which, in the $\mathcal{H}_{app}$-representation, corresponds to the \emph{diagonal} Hermitian operator $F = \sum_i f_i \ket{x_i} \bra{x_i}$, where $f_i \in \mathbb{R}$ is the measured value of the observable $f$ for the apparatus in state $\ket{x_i}$\footnote{This identity, presenting the observable $F$ as a linear combination of projectors, highlights how in the classical model the ontology of the observable is not distinguished from that of the state, as the state itself is expressed in logical-operational terms by its own projector $\ket{x_i}\bra{x_i}$. Note also that $f_i = f(x_i) = \bra{x_i} F \ket{x_i}$ as in the quantum case.}. Since all observables (and the projectors representing states) must be represented by diagonal operators in the fixed pointer basis, they all commute. Therefore, the experimental propositions represented by these projectors are all mutually compatible and constitute the ordinary Boolean logic that describes the classical apparatus.

\subsection{Causal duality: unitary or indeterministic}

Within the quantum model, according to the Copenhagen interpretation, two possible ways of physical evolution of the state of the object of inquiry are postulated: one \emph{unitary}, for the observed but not yet measured system (von Neumann process 2); the other \emph{indeterministic}, for the measured system (von Neumann process 1)~\cite{vneumann2}. The unitary evolution is governed by the \emph{Schr\"odinger equation} which, given the system \emph{Hamiltonian} and a fixed initial state, causally determines the state at any subsequent instant. The indeterministic evolution, on the other hand, is governed by the \emph{reduction postulate} by fixing the state, upon measurement completion, in an eigenstate of the measured observable\footnote{Von Neumann justifies this simplifying assumption by supposing a physical-analytical continuity in the measurement process; that is, the repetition of the measurement at an instant immediately following its completion will necessarily produce the same result, in accordance with the eigenstate-eigenvalue link. This basically constitutes the operative definition of measurement completion.}. Therefore, in the measurement, the unitary evolution of the overall system primarily concerns the \emph{physical} interaction between the apparatus and the object during the observation phase. The reduction process, however, refers to this same interaction at the moment of its completion by the analytical measurer -- that is, when the result is \emph{analytically} determined\footnote{It should be emphasized that, contrary to what is often understood, the assignment of a measurement outcome does not necessarily involve ``consciousness''. Generally, it is a physical act, analytically appropriate, that does not, however, imply a ``psychic nature'' on the part of the one performing it. It is true, though, that the human observer possesses an analytical nature.}. The principle of physical-analytical synergy ensures that both modes of evolution are co-present without contradiction. Thus, from the dual perspective, the so-called ``measurement problem''~\cite{measprob} -- for which the coexistence of the two modes appears unjustified and unclear -- is dissolved.

The \emph{causal duality} described above stems directly from the \emph{epistemic duality}, which requires a distinction between the models of the apparatus and the object. Therefore, the universalist perspective, according to which observed reality must be describable exclusively by means of the quantum model, cannot be accepted. Only an appropriate distinction between observation and measurement enables the understanding of the reduction process as physical-analytical, thereby balancing the nature of the observer with that of the measured object. In this way, the dual state evolution formally reflects its \emph{ontic} meaning -- a description of the \emph{measured} physical reality -- while, through the attribution of a value by the measured observable, it simultaneously acquires an \emph{epistemic} meaning, in accordance with the \emph{observing} analytical reality\footnote{It is worth clarifying the meaning of this duality in temporal evolution for a \emph{purely classical} system, such as an apparatus and its \emph{classical} measured system. In this context, no reduction process occurs upon measurement completion, as there is no ontic and epistemic distinction between observing and measuring. Consequently, the overall system undergoes no evolution in its classical Hilbert state space.}.

\subsection{Information duality: compatible or complementary}

The ultimate conceptual level of duality is found in the unity and distinction between \emph{compatible} and \emph{complementary} observables~\cite{complement}. The former can simultaneously form a descriptive basis for the state of the measured system; the latter, within the given measurement logos, ensures an \emph{exhaustive} experimental description of the object of inquiry in its various \emph{tropoi} of existence\footnote{When applied to the incompatible observables of momentum and position, this duality is historically expressed as ``wave-particle duality'', a hallmark of the early Copenhagen interpretation.}.

If the complete set of compatible observables adopted in the description of the phenomenon includes the Hamiltonian of the measured system, and its initial state is a Hamiltonian eigenstate, the reduction process of the state upon a later measurement does not imply a discontinuity with respect to the unitary evolution. The initial state, appropriately prepared (and therefore measured), is \emph{stationary} and persists over time -- up to a global phase -- in its own tropos of actuality\footnote{The stationarity of the object state represents the actuality of its existential tropos; on the other hand, the stationarity of trajectories in phase space represents the actuality of the existential tropos of the apparatus.}. This constitutes the \emph{permanent knowledge} that the observer has of the object, as enacted by the measurer upon the completion of the preparation. Consequently, observation and measurement are brought together in a single act of definitive mastery, and the experimental context defined by this set of compatible observables can be called \emph{essential}. It corresponds to the shared measurement logos, and \emph{the Hamiltonian represents the objective ontological essence of the object of inquiry}. Conversely, if the Hamiltonian is \emph{complementary} with the complete set of compatible observables, the provided description is \emph{non-essential}. That is, the initial state, as prepared in such a context, does not persist over time due to non-stationary unitary evolution; and the knowledge constituted by the preparation measurement generally degrades until a new measurement updates it through the reduction process. The observation period occurring between the completion of the preparation of the initial state and that of the subsequent measurement -- or more generally, between two measurements -- can be termed the \emph{information} phase.

The information phase is generally characterized by physical indeterminacy and analytical ambiguity. This condition persists throughout the observation, until the measurer resolves it by establishing physical knowledge of the measured object and analytical knowledge for the observer. This indeterminacy and ambiguity are \emph{intrinsic} and eminently \emph{probabilistic}, quantifying the statistical transitions from potentiality to actuality brought about by repeated measurements. Given that the potentiality-actuality transition is \emph{existential}, \emph{quantum probability} shares this same feature. This is not merely epistemic uncertainty about an existence that is already actualized, as in the classical setting; rather, it is an \emph{ontological} indeterminacy that can be resolved only through the measurement completion. Therefore, the \emph{quantum information} quantifies the physical-analytical knowledge that the act of measurement can provide relative to the pre-established measurement logos. Thus a quantum system, while measured, is informatively \emph{open} to knowledge. It is, instead, informatively \emph{closed} to knowledge and evolves unitarily when information remains constant and fixed\footnote{For this reason, a closed quantum universe can observe itself, but not properly measure itself, except within an \emph{essential} context devoid of information flow; that is, within a definitive context of \emph{objective} and actual existence.}. 

\section{The Copenhagen formalism of measurement}
\label{measure}

A fundamental problem of Quantum Mechanics, and of Physics in general, is to provide an adequate formalism for \emph{measurement}. In the Copenhagen interpretation, two stages are distinguished in the measurement process: the physical interaction between the system and the apparatus, and its analytical completion as determined by the measurer in conformity with the measurement logos. In the following, we describe how these stages are formalized by the theory, with the constraint of reserving the classical model for the apparatus and the quantum model for the system to be measured. The Copenhagen framework, in fact, is expressly \emph{not} quantum universalist: \emph{the classicality of the apparatus is crucial to account for the analyticity of the observer in the physical description of the observed reality}.

\subsection{Classical Hilbertian dynamics}

As discussed above, the classical description of the apparatus $A$ can be formulated using a Hilbert-space formalism for the pointer states, in order to make it compatible with the quantum description of the object to be measured. Let $\mathcal{H}_{A}$ represent the Hilbert space of the classical states of the system; then each state can be described as a basis vector $\ket{E_i}$, where $E_i$ represents the \emph{energy} assigned to the phase-space trajectory that characterizes it. If the state of the system is not known, but only the (\emph{epistemic}) probability $p_i > 0$ that it is in the state $\ket{E_i}$, the description can be made by means of the classical density matrix $\rho_A = \sum_i p_i \ket{E_i} \bra{E_i}$, which is diagonal in the fixed orthonormal classical basis, with $\sum_i p_i =1$\footnote{The mean value of any classical observable $F=\sum_j f_j \ket{E_j} \bra{E_j}$ in the state $\rho_A$ is then given by $\langle F \rangle = \sum_j p_j f_j = \mathrm{Tr} (F \rho_A)$.}. This means that a kind of expressly epistemic \emph{analytical} decoherence is imposed to account for observational activity. Any temporal evolution of the state of the system, which constitutes the classical information of the observer, implies an update to the density matrix at time $t$, $\rho_A(t) = \sum_i p_i(t) \ket{E_i} \bra{E_i}$, involving the probabilities $p_i$. In this way, the projector $\mathbb{P}_A = \sum_i \ket{E_i} \bra{E_i}$, associated with the experimental proposition that ``the system $A$ is in one of the states $\ket{E_i}$'', remains constant over time. 

In the general case, the temporal evolution of the state $\rho_A$ is described by a classical Markov process governed by a stochastic transition matrix $T(t)$ that describes the state transition probabilities~\cite{markov}. The following relation holds 
\begin{equation}
\rho_A(t) = \Lambda(t) [\rho_A (0)] = \sum_j K_j(t) \rho_A (0) K_j^\dagger(t) ,
\end{equation} 
where $K_j(t)$ are the \emph{classical} Kraus operators~\cite{kraus} defined as 
\begin{equation}
K_j(t) = \sum_i \sqrt{T_{ji}(t)} \ket{E_j} \bra{E_i}. 
\end{equation}
The elements of the stochastic matrix $T_{ji} \geq 0$ represent the transition probabilities from the state $\ket{E_i}$ to the state $\ket{E_j}$ at time $t$, and are such that $\sum_j T_{ji} (t) = 1 \quad \forall i$, so that $\mathrm{Tr} [ {\rho(t)} ] = 1$. Then, $T(t)$ can be expanded for an infinitesimal time $dt$:
\begin{equation}
T(dt) \simeq I + G dt + o(dt^2),
\end{equation}
where $G$ is the infinitesimal generator of the Markov process with the properties $ G_{ji} \ge 0$, $G_{ii} = -\sum_{j\neq i} G_{ji}$, and $\sum_j G_{ji} =0$ for all $i,j$. Therefore, the evolution equation for the probabilities $p_i(t)$ of the statistical mixture will be
\begin{equation}
\frac{dp_j(t)}{dt} = \sum_i G_{ji} p_i (t),
\end{equation}
which, in terms of the density matrix, becomes the \emph{master equation}\footnote{The equation is a special form of the Lindblad equation for open quantum systems (interacting with an environment)~\cite{lind}, constrained to the classical case where the Hamiltonian and the density matrix are diagonal. In our case, however, the classical system can also be closed.}
\begin{equation}
\frac{d \rho_A}{dt} = \sum_{i,j} G_{ji} \bra{E_i} \rho_A \ket{E_i} \ket{E_j} \bra{E_j}. 
\label{master}
\end{equation}
Equation (\ref{master}) has the following fundamental properties, which are necessary for the model to be well defined:
\begin{itemize}
\item the evolution is linear in $\rho_A$, since $\frac{d \rho_A}{dt}$ depends linearly on the probabilities $p_i$;
\item the trace of $\rho_A$ is preserved during the evolution, since $\mathrm{Tr}\left(\frac{d \rho_A}{dt}\right) = \sum_{i,j} G_{ji} p_i = \sum_i p_i \left(\sum_{j} G_{ji}\right) = 0$;
\item the diagonality of $\rho_A$ is preserved during the evolution, since $\rho_A(0)$ is diagonal, only diagonal elements are involved in subsequent instants;
\item the probabilities $p_i$ remain non-negative, given that $G_{ji} \geq 0$ for $j \neq i$.
\end{itemize}
Unlike in the quantum case, even for a closed classical system, the evolution of its state $\rho_A(t)$ is not generally unitary but stochastic\footnote{This consideration is the foundation of some realist interpretations of Quantum Mechanics, which hold that the Hamiltonian description via the Schr\"odinger equation is only an approximation of a stochastic formalism~\cite{ghirardi}.}. Consequently, the Hamiltonian does not appear explicitly in the master equation; instead, it influences the dynamics only indirectly through the probability distribution $p_i$, which depends on the energies $E_i$. Classical information is conserved only when $\rho_A(t)$ remains constant over time -- that is, when the condition of \emph{detailed balance} for state transitions is met. This indicates that, for the classical system, information has a quite different connotation than its quantum counterpart, taking on an exclusively epistemic meaning.

\subsection{Hybrid measurement states}

To include the object of inquiry $S$, in addition to the classical apparatus, in the description of the temporal evolution, substantial modifications are necessary with respect to the previously presented formulation. Indeed, the overall system cannot be considered purely classical, governed by the master equation (\ref{master}). Rather, it is a closed overall \emph{hybrid} system~\cite{hyb} whose state must be described by properly taking into account the different descriptions of the apparatus and the object of inquiry\footnote{The apparatus $A$ can, in a sense, be considered the classical physical ``environment'' for the measured system $S$.}.

Let $\mathcal{H}_S$ be the Hilbert space of the states of system $S$, which we can assume, for simplicity, to be finite-dimensional. The quantum model allows for the state of $S$ to be described by the quantum density matrix given by $\rho_S = \sum_{k,l} \rho_{kl} \ket{\phi_k}\bra{\phi_l}$, if $\{\ket{\phi_k}\}$ is an orthonormal basis of $\mathcal{H}_S$ consisting of common eigenvectors of a complete set of compatible observables for $S$, as defined by the experimental context. Thus, for $S$, the density matrix does not necessarily have to be diagonal. The description of the states of the overall system, formed by the apparatus and the object of inquiry, is performed via the usual tensor product space $\mathcal{H}=\mathcal{H}_A\otimes \mathcal{H}_S$, which constitutes the Hilbert space of the overall quantum-classical states. If $\rho_{AS}$ represents a \emph{hybrid} measurement state on $\mathcal{H}$, described by a density matrix, it must satisfy the reduction relation\footnote{The term ``reduction'', referring here to the description of a subsystem derived from the larger system that encompasses it, is not to be confused with the term in the process of wave function reduction.}
\begin{equation}
\mathrm{Tr}_S(\rho_{AS}) \equiv \rho_A = \sum_i p_i \ket{E_i}\bra{E_i}.
\label{rida}
\end{equation}
Furthermore, it must possess the properties of Hermiticity, unit trace, and positivity. One can see (appendix \ref{a1}) that the condition (\ref{rida}) imposes the following form on the state $\rho_{AS}$:
\begin{equation}
\rho_{AS} = \sum_i p_i \ket{E_i} \bra{E_i} \otimes \rho_S^{(i)}, \label{mist}
\end{equation}
where $\rho_S^{(i)}$ represents the quantum \emph{density matrix conditioned on the state} $\ket{E_i}$ and satisfies the reduction condition
\begin{equation}
\mathrm{Tr}_A (\rho_{AS}) \equiv \rho_S =  \sum_i p_i \rho_S^{(i)}.
\label{rids}
\end{equation}
Therefore, the classicality of $A$ requires that the measurement state $\rho_{AS}$, represented in the basis $\{ \ket{E_i} \otimes \ket{\phi_k} \}$, has a particular block structure such that there are no physical coherences between different states of $A$ and that the allowed correlations between $A$ and $S$ are exclusively classical, with no quantum entanglement\footnote{A universalist approach to Quantum Mechanics posits that the apparatus itself, similarly to the object of inquiry, is described quantum mechanically and that, due to physical decoherence, its state reduces to a diagonal form in a very short time~\cite{decoer1, decoer2}. In the Copenhagen interpretation, the apparatus cannot be described quantistically as an object of inquiry since this would imply another experimental context. The diagonal form of the apparatus state is instead assumed a priori, as a reflection of the analytical nature of the human observer and the classical mindset that designed it, that is unable to recognize superpositions.}. We also note that the form of $\rho_{AS}$ given in (\ref{mist}) is unique, since, once the probabilities $p_i > 0$ are fixed by
\begin{equation}
p_i = \mathrm{Tr} [ (\ket{E_i}\bra{E_i} \otimes \mathbb{1}_S) \rho_{AS}],
\label{probp}
\end{equation}
the conditional density matrices $\rho_{S}^{(i)}$ are uniquely determined as
\begin{equation}
\rho_{S}^{(i)} = \frac{1}{p_i}\mathrm{Tr}_A[(\ket{E_i}\bra{E_i} \otimes \mathbb{1}_S) \rho_{AS} ].
\end{equation}
In this way, the observation of the classical apparatus in the state $\ket{E_i} \bra{E_i}$, which resolves the epistemic uncertainty represented by $\rho_A$ by setting $p_i=1$, also imposes the related quantum state $\rho_S^{(i)}$ for the system $S$. However, unlike the apparatus, $\rho_S^{(i)}$ does not necessarily correspond to an actual existence for the object of inquiry (that is, to an eigenstate of the measured contextual observable) until the measurement completion; this crucial distinction expresses the \emph{essential ontological asymmetry} predicted by the Copenhagen interpretation~\cite{asym}.

\subsection{Measurement interaction}

The overall system, consisting of the apparatus ($A$) and the object of inquiry ($S$), is a closed physical universe during the interaction phase between the two subsystems within the process of measurement. This requires the conservation of information, which can be guaranteed by a unitary evolution driven by a suitable Hamiltonian of measurement $H_{AS}$. Indeed, even if the apparatus is described classically, the presence of the quantum system $S$ allows for such a Hamiltonian treatment of the overall system, as if it were purely quantum. In any event, the Hamiltonian $H_{AS}$ must be such that \emph{the classicality of the apparatus is rigorously preserved during the evolution}. This is equivalent to saying that \emph{the reduced density matrix $\rho_{A} (t)$ remains diagonal in the fixed basis $\{ \ket{E_i} \} $ throughout the interaction phase, and no coherences are generated between different states of $A$}. The classicality condition, therefore, imposes constraints on the form of the Hamiltonian $H_{AS}$ on $\mathcal{H}_A \otimes \mathcal{H}_S$ that we intend to specify.  

The essential ontological asymmetry, according to which the system $S$ cannot alter the actual existential tropos of the apparatus $A$ during the measurement interaction, requires that the (classical) Hamiltonian of the apparatus $H_A= \sum_i E_i \ket{E_i} \bra{E_i} $ on $\mathcal{H}_A$, considered as an observable on the measurement state space $\mathcal{H}_A \otimes \mathcal{H}_S$, is \emph{compatible} with $H_{AS}$; whereas this is not generally true for the free Hamiltonian of the system $H_S$ on $\mathcal{H}_S$\footnote{The compatibility of $H_A$ with $H_{AS}$ means that $H_A$ belongs to the \emph{essential} measurement context; i.e., at every moment of the measurement interaction, contrary to the case of the system $S$, the apparatus $A$ realizes its logos as expressed by the observer and shared by the measurer.}. For the compatibility of $H_A$ with $H_{AS}$, it is sufficient that the following commutation condition holds:
\begin{equation} [ H_{AS}, \ket{E_i} \bra{E_i} \otimes \mathbb{1}_S ] = 0. 
\label{comm}
\end{equation}
If, as usual, we decompose $H_{AS}$ as
\begin{equation}
H_{AS}= H_A \otimes \mathbb{1}_S+ \mathbb{1}_A \otimes H_S + H_{mis},
\end{equation}
the commutation condition given in (\ref{comm}) reduces to the one on the Hamiltonian term describing the measurement interaction, $H_{mis}$:
\begin{equation}
[ H_{mis}, \ket{E_i} \bra{E_i} \otimes \mathbb{1}_S ] = 0;
\label{comm1}
\end{equation}
and a form of $H_{mis}$ that satisfies the commutation relation (\ref{comm1}) is
\begin{equation}
H_{mis} = \sum_i \ket{E_i} \bra{E_i} \otimes V_i,
\end{equation}
where $V_i$ are Hermitian operators defined on $\mathcal{H}_S$ which describe the measurement interaction depending on the state $\ket{E_i}$ of $A$. Therefore, $H_{AS}$ can be ultimately written as
\begin{equation}
H_{AS} = \sum_i E_i \ket{E_i} \bra{E_i} \otimes \mathbb{1}_S + \mathbb{1}_A \otimes H_S + \sum_i \ket{E_i} \bra{E_i} \otimes V_i ,
\label{has}
\end{equation}
and such a form of $H_{AS}$ is the \emph{unique} one compatible with the classicality of the apparatus. It is, therefore, worth examining the motivations for this.

To prove that the form in (\ref{has}) preserves the classicality of the apparatus during the measurement interaction, let us assume that the composite system is, according to relation (\ref{mist}), in the initial state \begin{equation} 
\rho_{AS} (0) = \sum_j p_j (0)\ket{E_j} \bra{E_j} \otimes \rho_S^{(j)} (0), 
\label{initr}
\end{equation} 
and that $\rho_A(0)$ is diagonal. Then, the unitary evolution of the state $\rho_{AS} (t)$ during the measurement interaction, governed by the von Neumann equation
\begin{equation}
\frac{d\rho_{AS}}{dt} = -\frac{i}{\hbar} [H_{AS}, \rho_{AS}],
\end{equation}
prescribes
\begin{equation}
  \rho_{AS} (t) = e^{-i H_{AS} t/ \hbar} \rho_{AS} (0)  e^{i H_{AS} t/ \hbar} .
\label{ras}
\end{equation}
Since $H_{AS}$ commutes with the projectors $\ket{E_j} \bra{E_j} \otimes \mathbb{1}_S$, the evolution preserves the block structure of $\rho_{AS} (0)$ given in (\ref{initr}) and, from (\ref{ras}),  we have (appendix \ref{a2})
\begin{equation}
  \rho_{AS} (t) = \sum_j p_j (0) \ket{E_j} \bra{E_j} \otimes \rho_S^{(j)} (t),
  \label{rast}
\end{equation}
where the conditional density matrices $\rho_S^{(j)} (t)$ satisfy the relation  
\begin{equation}
\rho_S^{(j)} (t) = e^{-i (H_S + V_j) t/ \hbar} \rho_S^{(j)} (0)  e^{i (H_S + V_j) t/ \hbar}.
\label{rhoit}
\end{equation}
Therefore, the von Neumann equation also holds for $\rho_S^{(j)}$, depending on $\ket{E_j}$:
\begin{equation}
\frac{d\rho_{S}^{(j)}}{dt} = -\frac{i}{\hbar} [H_{S}+V_j, \rho_{S}^{(j)}],
\label{vneumann}
\end{equation}
and, like the overall system, $\rho_S^{(j)}$ evolves unitarily, but under the effective Hamiltonian $H_S + V_j$. Instead, the state of the apparatus during the measurement interaction is given by
\begin{equation}
\rho_A (t) = \mathrm{Tr}_S [\rho_{AS}(t)] = \sum_i p_i (0) \ket{E_i} \bra{E_i},
\label{astatet}
\end{equation}
which is diagonal and constant in time, thus respecting classicality\footnote{The constancy of the state of the apparatus and of the epistemic probabilities $p_i$ means that the associated classical information is conserved and that, during the measurement interaction, the detailed balance condition holds for the transitions between the states $\ket{E_i}$. In other words, the classical apparatus is a closed system during the measurement interaction.}.

In addition to being sufficient for the classicality of the apparatus, the form of $H_{AS}$ in (\ref{has}) is also necessary. To prove this, we consider the general form
\begin{equation}
 H_{AS} = \sum_{i,j} \ket{E_i} \bra{E_j} \otimes W_{ij}
 \label{hasgen}
\end{equation}
and assume that an off-diagonal coupling is allowed, i.e., between distinct states of $A$. This means that, for some pair $k, l$ with $k \neq l$, we have $W_{kl} \neq 0$. We evaluate the matrix element $ \bra{E_k} \rho_A (t) \ket{E_l}$ for short times after the initial instant $t=0$ of the evolution of the measurement system. Setting, for simplicity, the initial epistemic condition $p_k(0) = 1$ and assuming that the system $S$ is initially prepared in a pure state $\rho_S^{(k)} (0) = \ket{\phi}\bra{\phi}$, from  the expansion $ e^{- i H_{AS} t / \hbar} \simeq \mathbb{1} - i H_{AS} t / \hbar + o(t^2)$ and
\begin{equation}
\begin{split}
\bra{E_k} \rho_A (t) \ket{E_l} & = \mathrm{Tr} \left[ (\ket{E_l} \bra{E_k}  \otimes \mathbb{1}_S) \rho_{AS} (t)  \right] \\
  & = \mathrm{Tr} \left[ (\ket{E_l}  \bra{E_k} \otimes \mathbb{1}_S) e^{- i H_{AS} t / \hbar} \rho_{AS} (0) e^{ i H_{AS} t / \hbar}  \right],
\end{split}
\label{expans}
\end{equation}
we obtain (appendix \ref{a3})
\begin{equation}
\bra{E_k} \rho_A (t) \ket{E_l} \simeq  \frac{it}{\hbar} \bra{\phi} W_{kl }\ket{\phi} + o(t^2) . 
\label{elemr}
\end{equation}
Since $W_{kl}\neq 0$, there certainly exists a state $\ket{\phi}$ such that $\bra{\phi} W_{kl }\ket{\phi} \neq 0$; then, the classicality of the apparatus is violated. 

Knowing the unitary evolution of the conditional density matrices $\rho_{S}^{(i)}$, as given by (\ref{vneumann}), the generally \emph{not} unitary evolution for the state $\rho_S$ is obtained from (\ref{rids}), recalling that the probabilities $p_i$ remain constant in time. Therefore, the apparatus $A$ acts as an external ``controller'' that modulates the evolution of the system $S$ through suitable measurement potentials $V_i$. However, in accordance with the essential ontological asymmetry, $A$ is in principle not influenced by $S$, and thus no coherences are created. Furthermore, there is no entanglement between $A$ and $S$, since there are no interaction terms that mix different states of $A$.

As an example, we can assume that $S$ is a qubit in the initial state $\ket{+} = \frac{\ket{0} + \ket{1}}{\sqrt{2}}$ and that the apparatus $A$ admits two pointer states for the measurement of the spin, $\ket{\textrm{spot-up}}$ and $\ket{\textrm{spot-down}}$, with $p_1=p_2=1/2$, in such a way that 
\begin{equation}
\rho_A = \frac{1}{2} \ket{\textrm{spot-up}} \bra{\textrm{spot-up}} + \frac{1}{2} \ket{\textrm{spot-down}} \bra{\textrm{spot-down}}.
\end{equation} 
Then, an interaction of the Stern-Gerlach type, with $H_S=0$, $V_1= g \sigma_z$ and $V_2=-g \sigma_z$, constitutes the measurement interaction. By setting $\omega= g/\hbar$, from (\ref{rhoit}) we have
\begin{equation}
\rho_S^{(1)} (t) = e^{-i \omega \sigma_z t } \ket{+}\bra{+} e^{i  \omega \sigma_z t } = \left( \begin{matrix} \cos^2 \omega t & -i \cos \omega t \sin\omega t \\ i \cos \omega t \sin \omega t & \sin^2 \omega t \end{matrix} \right)
\end{equation}
and
\begin{equation}
\rho_S^{(2)} (t) = e^{i \omega \sigma_z t} \ket{+}\bra{+} e^{-i \omega \sigma_z t } = \left( \begin{matrix} \cos^2 \omega t & i \cos \omega t \sin \omega t \\ -i \cos \omega t \sin \omega t & \sin^2 \omega t \end{matrix} \right),
\end{equation}
expressed in the basis $\{\ket{+}, \ket{-}\}$, where $\ket{+} = \frac{\ket{0} + \ket{1}}{\sqrt{2}}$ and $\ket{-}= \frac{\ket{0} - \ket{1}}{\sqrt{2}}$. Therefore, from (\ref{rids}), the state of $S$ during the measurement interaction will be
\begin{equation}
\rho_S(t) = p_1 \rho_S^{(1)} (t) + p_2 \rho_S^{(2)} (t) = \left( \begin{matrix} \cos^2 \omega t & 0 \\ 0 & \sin^2 \omega t \end{matrix}\right),
\end{equation}
i.e., a statistical mixture of $\ket{+} \bra{+}$ and $\ket{-} \bra{-}$ with weight probabilities $\cos^2 \omega t$ and $\sin^2 \omega t$, respectively. Note that, as a consequence of the principle of physical-analytical synergy, the analytical decoherence of apparatus $A$ automatically translates into a physical decoherence in system $S$ without the need for additional mechanisms. This may suggest that the phenomenon of physical decoherence is closely related to the mobility of the cut; and it also highlights the physical-analytical connotation of the notion of a measured state, which is expressed through the epistemic meaning of classical probabilities and the ontic meaning of quantum ones.

\subsection{Measurement completion}

The mere interaction of a system, named ``apparatus'', with a quantum system does not generally constitute a measurement, despite the induced correlations and certain knowledge of its classical state from a given instant onwards. For the interaction to effectively constitute a measurement, its \emph{completion} is necessary: that is, \emph{the moment when the information about the measured system is transformed into knowledge by that very same act, in accordance with the measurement logos}. Then, the measurement completion admits a dual perspective: 
\begin{itemize}
\item from the point of view of the apparatus $A$, as observed by the observer, the \emph{epistemic} uncertainty related to \emph{classical} information
\begin{equation}
I_A = -K \mathrm{Tr} (\rho_A \ln \rho_A)
\label{classici}
\end{equation}
is discontinuously transformed into certainty ($I_A=0$) upon the \emph{registration} of a specific pointer state $\ket{E_l}$\footnote{Since $\rho_A$ is diagonal, $I_A$ is the usual Shannon information (or \emph{entropy}), where $K$ represents a suitable constant~\cite{sh1}. For example, in a thermodynamic context, it may represent Boltzmann's constant.}; 
\item from the point of view of the system $S$, as measured by the measurer, the \emph{ontic} indeterminacy related to \emph{quantum} information
\begin{equation}
I_S = -K \mathrm{Tr} (\rho_S \ln \rho_S)
\label{quantumi}
\end{equation}
is discontinuously transformed into determination ($I_S=0$) upon the \emph{reduction} of the object of inquiry to a particular state $\ket{\phi_m}$ within the predetermined experimental context\footnote{The information $I_S$ is generally hybrid, with its value depending on the epistemic weights $p_i$, leading to temporal variation. It is purely ontic and time-invariant only under the condition of epistemic certainty regarding the state of the apparatus.}.
\end{itemize}
Consequently, the information of the physical universe $I_{AS}= -K \mathrm{Tr} (\rho_{AS} \ln\rho_{AS})$, which remains constant in the measurement interaction, discontinuously changes to $I_{AS}=0$ upon measurement completion\footnote{This circumstance seems to indicate the ``arrow of time'' in the sequence of measurement completions that progressively erase quantum information. This suggests a fundamentally analytical structure of time, dependent on the observer-measurer system, that is not continuous. Even some interesting theoretical models also admit a discontinuous ``granularity'' for time~\cite{time}.}.

The physical recording of the ``measurement result'' by the apparatus represents the epistemic foundation of measurement. Recalling (\ref{rida}), it occurs with a classical probability 
\begin{equation} p_l = \mathrm{Tr} (\ket{E_l} \bra{E_l} \rho_A) ,
\end{equation} 
and reveals the apparatus $A$ in the definite state $\ket{E_l}$. Thus, upon registration, the state of the system $S$, $\rho_S$, is revealed to be $\rho_S^{(l)}$, and subsequently it evolves unitarily, preserving the associated quantum information as given by (\ref{quantumi}). In other words, the physical recording \emph{closes} the system $S$ to knowledge, awaiting the measurement completion by an analytical act of the measurer.

The knowledge of the measured object is ontically realized by the measurer in the reduction of its state, $\rho_S^{(l)}$, which is determined upon measurement completion. By the Born rule, this reduction occurs with a conditional quantum probability 
\begin{equation} \pi_m^{(l)} = \mathrm{Tr} [ \ket{\phi_m} \bra{\phi_m} \rho_S^{(l)}] ,
\end{equation} 
and leaves it in the new state $\ket{\phi_m}$, which is an eigenstate of the measured observable. If the measured observable is \emph{essential}, i.e., if it is compatible with the measurement Hamiltonian, the knowledge acquired upon the completion of the measurement persists over time (and  $I_S =0$ constantly). If it is \emph{non-essential}, the state of $S$ is prepared by the measurement, but the system information evolves over time until the subsequent registration by the measurer. Then, the measurement completion finally translates the information, fixed at the registration, into new objective and shareable knowledge.

\section{Key paradoxes in quantum measurement}
\label{paradoxes}

Erwin Schr\"odinger presented, probably not without a subtle polemical intent, what would become the most famous paradox of Quantum Mechanics in a celebrated 1935 article, published in three parts in \textit{Naturwissenschaften}~\cite{cat}. Accordingly, a cat enclosed in a steel chamber would be in the state ``alive and dead'' as a reflection of its correlation with the quantum indeterminacy of an atomic decay that triggers an evil poison release. Does opening the chamber \emph{reveal} or, rather, \emph{determine} that the cat is alive or dead? About twenty-six years later, in 1961, Eugene Wigner proposed a refinement of Schr\"odinger's cat thought experiment~\cite{friend}, in which a human ``friend'' -- a physicist sometimes playfully identified with his colleague and countryman von Neumann -- rather than the cat, can be observed while conducting the atomic decay experiment (fortunately without any release of poison). In this case, the correlation between the friend and the quantum indeterminacy associated with the decay would seem to implicate consciousness in the reduction of the atom wave function. To our knowledge, none of the exponents of the Copenhagen School ever publicly and directly commented on the two proposed scenarios. In this section, we attempt to provide a possible response in light of the considerations developed in the previous sections.

\subsection{Schr\"odinger's cat paradox}

Here is Trimmer's translation of the passage that launched ``Schr\"odinger's cat'' into History of Physics and Philosophy of Science~\cite{cat}.
\begin{displayquote}
One can even set up quite ridiculous cases. A cat is penned up in a steel chamber, along with the following diabolical device (which must be secured against direct interference by the cat): in a Geiger counter there is a tiny bit of radioactive substance, \emph{so} small, that \emph{perhaps} in the course of one hour one of the atoms decays, but also, with equal probability, perhaps none; if it happens, the counter tube discharges and through a relay releases a hammer which shatters a small flask of hydrocyanic acid. If one has left this entire system to itself for an hour, one would say that the cat still lives \emph{if} meanwhile no atom has decayed. The first atomic decay would have poisoned it. The $\psi$-function of the entire system would express this by having in it the living and the dead cat (pardon the expression) mixed or smeared out in equal parts. 
\end{displayquote}
More than with the measurement problem in itself, and the meaning of the state reduction, here Schr\"odinger seems primarily interested in the description that the quantum model provides of the state of the overall system as ``having in it the living and the dead cat [...] mixed or smeared out in equal parts''. This description, in fact, contrasts with the ordinary classical scheme, according to which the ``cat'', as a macroscopic object, is simply locked in the chamber ``alive'' \emph{or} ``dead'' with equal probability. To appropriately address the issue, it is first necessary to focus on the analytical and epistemological dimensions of the duality, and then identify the experimental context by positioning the Heisenberg cut. The ambiguity pointed out by Schr\"odinger, in fact, lies in the universalist approach adopted, which intends to generically describe the overall system without specifying the experiment in which it becomes an object of inquiry and, therefore, quantum. In the absence of this specification, the system, including the cat, cannot be described quantum mechanically. Instead, the passage highlights that the role of the cat is simply that of a mere ``detector'': the vital state described by the \emph{formal} attributes ``alive'' or ``dead'', proper to the classical mindset of the observer, is correlated with the quantum state of the atom, described by means of the \emph{context-natural} attributes ``decayed'' or ``non-decayed''. Therefore, the description pertaining to the cat \emph{must be classical}; only that of the atom is properly quantum.

The logic of experimental propositions applicable to the object ``cat'', for what has been said, is ordinarily Boolean. In the Hilbert space of its states, the orthonormal basis $\{ \ket{\mathrm{alive}}, \ket{\mathrm{dead}} \}$ is fixed, and the operator $\ket{\mathrm{alive}} \bra{\mathrm{alive}} + \ket{\mathrm{dead}} \bra{\mathrm{dead}} = \mathbb{1}_\mathrm{cat} $ is associated with the true experimental proposition ``the cat is alive \emph{or} dead''. Note that, if $\ket{\mathrm{alive}}$ and $\ket{\mathrm{dead}}$ are unambiguous orthogonal states, the experimental proposition ``the cat is alive \emph{and} dead'' is false in any case, both in classical and quantum logic; indeed, the corresponding operator is $\ket{\mathrm{alive}} \bra{\mathrm{alive}}  \ket{\mathrm{dead}} \bra{\mathrm{dead}}=\mathbb{0}_{\mathrm{cat}}$. 
In analogy with (\ref{rida}), the uncertain state of the cat is then described, in principle, throughout the entire experiment by the classical statistical mixture
\begin{equation} 
\rho_{\mathrm{cat}} = \frac{1}{2} \ket{\textrm{alive}} \bra{\textrm{alive}} + \frac{1}{2} \ket{\mathrm{dead}} \bra{\mathrm{dead}},
\label{catstate}
\end{equation}
whose epistemic weights are set to $p_1=p_2=1/2$ by the experimental conditions. Then, from (\ref{rids}), the state of the atom is
\begin{equation}
\rho_\mathrm{atom} (t) = \frac{1}{2} \rho_\textrm{atom}^{\textrm{non-decayed}} (t) + \frac{1}{2} \rho_\mathrm{atom}^{\mathrm{decayed}} (t).
\label{atomstate}
\end{equation} 
In order to detect the quantum effects on the macroscopic ``cat'' system, it is necessary to design an explicit \emph{new} experimental context that highlights this \emph{new} phenomenon. This means that the cat transitions from being a classical ``detector'' of atomic decay to an object of quantum investigation. Such a context cannot be assumed to exist a priori, and it is fundamental to the description of the measurement state. In particular, one would need to answer the question: what is the experimental apparatus designed to detect the quantum effects on the cat as a macroscopic object? In other words, \emph{the formal attributes ``alive'' and ``dead'', although permitted in a classical description, must acquire a context-natural meaning in order to be used in the description of the quantum state of the cat}. Only under these conditions is it possible to consider the state of the cat as an entangled quantum superposition rather than the classical statistical mixture given in (\ref{catstate}); and this state of superposition would, over time, in any case, be changed into a quantum statistical mixture by the epistemic uncertainty about the state of the apparatus in the measurement interaction. In this way, the Copenhagen interpretation reduces the transition from superposition to a quantum statistical mixture to the analytical uncertainty about the states of the classical apparatus, and, synergically, to physical decoherence~\cite{maccone}.

Addressing the issue of whether opening the chamber reveals or, rather, determines that the cat is alive or dead involves the causal and information dimensions of the duality. The state of the cat, given in (\ref{catstate}), is stationary and, from (\ref{classici}), implies classical information, $I_{\mathrm{cat}} = K \ln 2$, which quantifies the epistemic uncertainty before opening the chamber while awaiting the measurement completion by the measurer, e.g., Schr\"odinger. In contrast to this time-invariant uncertainty, the indeterminacy of the (reduced) atomic state in (\ref{atomstate}), associated with its quantum information $I_\mathrm{atom}$ given in (\ref{quantumi}), varies in time according to the non-unitary evolution imposed by the overall measurement Hamiltonian. Once the chamber is opened, the epistemic uncertainty is dissolved and the state of the cat is \emph{revealed} for what it is (as well as that of the rest of the classical apparatus), ``alive'' or ``dead''. The manifestation of the cat state corresponds to a registration by the measurer that follows the physical recording by the measuring apparatus and immediately fixes the quantum information associated with the state of the atom. But \emph{only upon measurement completion by the measurer -- which drastically nullifies the ontic indeterminacy expressed by $I_\mathrm{atom}$ -- is this state analytically, as well as physically, determined and known}.

\subsection{Wigner's friend paradox}

Wigner imagines a scenario analogous to that of Schr\"odinger's cat, but it can be assumed that it is not the cat's unaware vital state that reveals the possible decay of the atom, but rather the conscious testimony of the physicist (the ``friend'') who conducted the decay experiment as a measurer. Therefore, the friend, enclosed in the chamber, enacts the standard description of the measurement by observing the decay-detecting apparatus and, then, measuring the atom in its state. The information about the possible decay is registered by the detecting apparatus and by the measurer's body; while the actual state of the atom is determined and known by the measurer upon measurement completion. In this experimental context, the friend's body constitutes a classical system whose state is uncertain for Wigner until the chamber is opened. This state can be described analogously to the cat state in (\ref{catstate}):
\begin{equation}
\rho_\mathrm{friend}  = \frac{1}{2} \ket{\textrm{no}} \bra{\textrm{no}} + \frac{1}{2} \ket{\mathrm{yes}} \bra{\mathrm{yes}},
\end{equation}
where $\ket{\mathrm{no}}$ and $\ket{\mathrm{yes}}$ represent physical states of the friend's body, alternatives to the cat states $\ket{\mathrm{alive}}$ and $\ket{\mathrm{dead}}$, indicating the response of the friend to the question ``has the atom decayed?''. Once the chamber is opened, Wigner, as a (master) observer who incorporates the measurer in the inquiring unity, is informed by his friend about the actual state of the atom. At that moment, the ontic information related to the atom \emph{is fixed and objectively confirmed as zero without discontinuity}. This is consistent with the measurement completion by the friend, which has \emph{already} occurred. There is therefore nothing new or paradoxical compared to the cat case, except for the fact that the measurement completion occurred while the chamber was closed\footnote{To use Wigner's own words~\cite{friend}: ``All this is quite satisfactory: the theory of measurement, direct or indirect, is logically consistent as long as I maintain my privileged position as ultimate observer''.}.

But let us now imagine a hypothetical experimental context in which the friend -- who is himself conducting the original experiment on the atom decay -- and the contents of the chamber are subjected to measurement by Wigner. This implies that the apparatus has been specifically designed to detect the quantum effects on the friend's body (that is, the friend's body is in turn also measured, not simply observed), and that the attributes ``no'' and ``yes'', corresponding to the friend's impressions about the outcome of the decay, are now context-natural rather than merely formal. In principle, in this context, assuming an initial superposition state for the atom, Wigner is justified in considering the system in an entangled superposition state that correlates the state of the atom with that of the friend's body, until he completes his measurement \emph{after} opening the chamber. In such a superposition state, the impressions produced on the friend's body by the outcome of the decay experiment do not correspond to definite properties; yet, the friend has already formed an awareness of the state of the atom \emph{before} the chamber is opened. Therefore, in this experimental context, the ``principle of psycho-physical parallelism'', as quoted in section \ref{intro}, that von Neumann assumes for the measurement theory, is contradicted; and the standard assumption that the atom is in a state of superposition upon the opening of the chamber renders the quantum description logically inconsistent. 

To preserve the ``principle of psycho-physical parallelism'' and the consistency of the theory, it is necessary that the state of the atom is already determined before Wigner opens the chamber; that is, that the consciously acquired impressions on the decay outcome correspond to definite properties of the friend's body. It seems, therefore, inevitable to link the measurement completion -- i.e., the reduction of the state of the measured system -- to the conscious acquisition of such impressions by the friend. Here is a final summary in Wigner's words~\cite{friend}: 
\begin{displayquote}
It follows that the being with a consciousness must have a different role in quantum mechanics than the inanimate measuring device [...]. In particular, the quantum mechanical equations of motion cannot be linear if the preceding argument is accepted. This argument implies that ``my friend'' has the same types of impressions and sensations as I -- in particular, that, after interacting with the object, he is not in that state of suspended animation which corresponds to the [entangled superposition state].
\end{displayquote} 
Wigner's conclusion is, however, unacceptable in the Copenhagen interpretation, since, according to it, the linearity of the theory is a constitutive requirement. Furthermore, the parts of the body of a conscious being are nothing special, being simply  links in the von Neumann chain, classical or quantum, depending on their position relative to the chosen physical-analytical cut. In any case, given the experimental context, the cut simply and clearly distinguishes between two \emph{physical tropoi}: the actual existence of the apparatus (with its environment) and the potential existence of the object; not a ``non-linear domain of the psyche'' and a ``linear domain of the physique''.

To address the fundamental issue raised by Wigner from the perspective of the Copenhagen interpretation, it is necessary to abandon the ``principle of psycho-physical parallelism'' and to specifically consider the ontological dimension of duality, the one that distinguishes between the complementary physical nature of the observed and the analytical nature of the observer, as well as between the measured, the measurer, and the observer. In fact, it is not so much the mathematical formalism (linearity or non-linearity) that is in question, but rather the logical consistency of the meaning of measurement, which requires that, \emph{within the experimental context properly designed to realize the measurement logos, a measurer cannot be simultaneously measured, but only observed}. This means that, \emph{the measurer is always unique: the first to complete the measurement}; while the other involved analytical components (whether explicit or not), who share the unique measurement logos and constitute the inquiring unity, complete the logical-communicative act that constitutes the observation. \emph{Only the measurer's act makes the existence of the measured object actual and objective for observation by definitively canceling the quantum information}.

What makes Wigner's scenario peculiar compared to Schr\"odinger's is the multiple analytical components that can share the measurement logos (Wigner and the friend) rather than just one (Schr\"odinger). In the latter case, the measurer is unequivocal and logical consistency is not threatened; whereas in the former case, to preserve logical consistency, it is necessary to clarify the roles of the analytical components; i.e., to properly locate the von Neumann cut between measured and measurer. Since Wigner and his friend share the same measurement logos, which requires the identification of the atomic decay outcome, it is the friend who, acting as the measurer, completes this measurement, determining the actual state of the atom. Thus, Wigner can only be a (master) observer, informed by his friend of the experimental outcome. The description of the measurement must take into account this \emph{logical} priority, regardless of the measurer's consciousness. 
 
\section{Conclusion}
\label{concl}

The Copenhagen interpretation of Quantum Mechanics is founded on duality. The multi-perspective schema adopted in this work proposes five distinct levels of duality which collectively define the Copenhagen philosophical criterion. These levels are encapsulated in a ``Russian doll'' structure, as shown in section \ref{dual}, and their essential characteristics are summarized below. 
\begin{enumerate}
\item \emph{Ontological duality}. A correspondence exists between the analytical nature of the observer, measurer, and measured, and the physical nature of the apparatus and object of inquiry. The von Neumann cut is the analytical distinction between the measurer -- incorporated into the observer -- and the measured, establishing their activities: the observer observes, the measurer is observed and measures, and the measured is observed and measured. Dually, the Heisenberg cut is the physical distinction between the apparatus -- embodied by the measurer -- and the object of inquiry, differentiating the observed apparatus from the measured object. The shared analytical measurement logos, physically manifested by the actual existence of the apparatus, also culminates in the actualization of the measured object's existence, thereby constituting the phenomenon.
\item \emph{Analytical duality}. Focusing on the analytical activity, specifically that of the human observers, a dual correspondence is established between formal language and context-natural language. While the former involves hypothetical attributes based on a given mindset, independent of the measurement, the latter requires the proper and explicit definition of the experimental context to generate the phenomenon-relative attributes. This tension is resolved through a threefold mode of disambiguation: first, by expressing the context; second, by localizing the cut to distinguish the measurer from the measured, and the apparatus from the object; and finally, by the physical-analytical act of measurement itself. 
\item \emph{Epistemological duality}. Within the framework of context-natural language, two models are provided for describing the phenomenon: the classical model, which pertains to the apparatus that physically constitutes the measurer; and the quantum model, which pertains to the physical object being measured. This distinction is not a symptom of an ``imperfect'' or ``incomplete'' theory, but rather the direct reflection of the ontological duality established by the cut. In fact, it expresses the distinction between the apparatus, which is properly definite in its actuality, and the object, which is physically-analytically ambiguous in its potentiality. Superposition is the watershed between the quantum and classical models, as it captures this ambiguity by formalizing the ontic distinction between the observable and the state. 
\item \emph{Causal duality}. The quantum model posits a dual mode of evolution for the state of the observed physical universe: unitary, when it has not yet been measured; and indeterministic, upon measurement completion. This duality mirrors the ontic distinction between the observed apparatus and the measured object, and makes explicit within the quantum model the distinction between observation and measurement, a distinction otherwise undetectable in a classical framework. In this way, the dual mode of evolution loses its ``problematic aspect'', formally encompassing the drastic transition from potentiality to actuality brought about exclusively by the measurement completion, alongside the unitarity of the mere observation.
\item \emph{Information duality}. The measurement completion aligns with the observation in only one case: when the experimental context is essential, that is, compatible with the Hamiltonian, and the state is stationary. Indeed, in this condition, measurement completion does not affect the stationarity, since the observed existence is already actualized, and the observer possesses physical-analytical knowledge of it without the need for information. The dual scenario involves a non-essential context, that is, complementary to the Hamiltonian, in which information is instead preparatory to physical-analytical knowledge. In this case, the information pertaining to the state generally varies over time until it is epistemically fixed by the apparatus's recording. It is then nullified as it is converted into physical-analytical knowledge upon measurement completion.
\end{enumerate}

The five levels of duality outlined above provide the criterion for identifying those interpretations of Quantum Mechanics that are genuinely inspired by the Copenhagen School, and they substantiate the principle of physical-analytical synergy formulated in Section \ref{intro}. In particular, the epistemological duality addresses the ``macro-micro'' dichotomy by defining the ``macro'' as the physical description that, unlike the ``micro'', admits no real distinction between state and observable, nor state superposition. On the other hand, the causal duality dissolves the notorious ``measurement problem'' by highlighting the epistemic distinction, within the quantum model, between  observation and measurement completion. Finally, the information duality resolves the meaning of physical-analytical knowledge in terms of the stationarity of the observed universe. 

To situate the epistemological duality within the analytical duality is the central aim of Bohr's work. He fundamentally regards this positioning as an exercise in logical disambiguation. For example, as argued in section \ref{paradoxes}, one can resolve the ``Schrödinger's cat'' and ``Wigner's friend'' paradoxes by eliminating ambiguity about the position of the cut. The ultimate objective, however, is to situate the analytical perspective relative to the ontological one. Bohr is prudent on this point -- a prudence perhaps often superficially misinterpreted as ``subjectivism''. Far more explicit is von Neumann, who grounds his measurement theory in the exclusive philosophical presupposition that the dual natural irreducibility of the observed and the observer represents the sole ontological possibility for ensuring the formal consistency of the physical description.

Despite his intuition of ontological duality, von Neumann, contrary to Bohr, seems to assume a universalist stance: the von Neumann chain formally regresses back to the observer, not stopping at the apparatus. In this way, the measurer becomes measured, not merely observed, making the measurement logos ambiguous. This is the crucial critique that the Copenhagen School poses to von Neumann's formalization. As shown in Section \ref{measure}, to address the problem and, nevertheless, respect the classicality of the measuring apparatus, its state space can be given a Hilbert space description, which, however, precludes superposition. This prescription renders the apparatus states decoherent in principle, but as an analytical reflection, not as a physical effect. Thus, physical decoherence in measured macroscopic systems is not confused with analytical decoherence in the apparatus. The movability of the cut -- that is, the freedom in choosing the experimental context -- explains how physical decoherence is the presupposition for analytical decoherence and justifies, through a bootstrapping process that connects the quantum physical reality to the classical description, the emergence of the classical mindset.  In this view, the observer is no longer a mere spectator of physical reality but an essential co-author of its very objective description.

\newpage

\appendix

\section{Supplement to the formalism of measurement}
In this appendix, we provide some operational insights into the measurement formalism, as outlined in section \ref{measure}.

\subsection{General form for $\rho_{AS}$}
\label{a1}

The general form of a quantum-classical state $\rho_{AS}$ is given by
\begin{equation}
\rho_{AS} = \sum_{i,j} \ket{E_i}\bra{E_j} \otimes \rho_{S}^{(ij)},
\end{equation}
where $\rho_{S}^{(ij)}$ represents a generic operator on $\mathcal{H}_S$. This can be decomposed as 
\begin{equation}
\rho_{AS} = \sum_i \ket{E_i} \bra{E_i} \otimes \rho_{S}^{(ii)} + \sum_{i \neq j} \ket{E_i} \bra{E_j} \otimes \rho_{S}^{(ij)},
\end{equation}
and, since $A$ is a classical system, no analytical coherences between different pointer states are allowed. This means that $\rho_{S}^{(ij)}=0$ for all $i \neq j$, and 
\begin{equation}
\rho_{AS} = \sum_i \ket{E_i} \bra{E_i} \otimes \rho_{S}^{(ii)}.
\end{equation}
The reduction relation then becomes
\begin{equation} 
\rho_A = \mathrm{Tr}_S(\rho_{AS}) = \mathrm{Tr}_S \left[ \sum_i \ket{E_i}\bra{E_i} \otimes \rho_{S}^{(ii)} \right],
\end{equation}
and, by linearity, we have
\begin{equation}
\rho_A = \sum_i \mathrm{Tr}_S \left[ \ket{E_i}\bra{E_i} \otimes \rho_{S}^{(ii)} \right] = \sum_i \ket{E_i}\bra{E_i} \mathrm{Tr} \left[ \rho_{S}^{(ii)} \right].
\end{equation}
The condition (\ref{rida}) can therefore be satisfied if, and only if, the following holds:
\begin{equation}
\mathrm{Tr} [\rho_{S}^{(ii)}] = p_i,
\label{trunity}
\end{equation}
and, by setting $\rho_S^{(i)} \equiv \frac{\rho_{S}^{(ii)}}{p_i}$ ($p_i > 0$), we obtain the form in (\ref{mist}).

For $\rho_S^{(i)}$ to effectively represent a conditional density matrix, it is necessary to verify the properties of Hermiticity, positivity, and unit trace. The Hermiticity of $\rho_S^{(i)}$ follows directly from that of $\rho_{AS}$, given that $\ket{E_i}\bra{E_i}$ is Hermitian and $p_i$ is real. To prove positivity, we consider a generic state $\ket{\phi} \in \mathcal{H}_S$ and the state $\ket{\Phi_i} = \ket{E_i} \otimes \ket{\phi} \in \mathcal{H}_A \otimes \mathcal{H}_S$. From positivity of $\rho_{AS}$, we have
\begin{equation}
\begin{split}
0 \leq \bra{\Phi_i} \rho_{AS} \ket{\Phi_i} & = (\bra{E_i} \otimes \bra{\phi}) \left[ \sum_j \ket{E_j}\bra{E_j} \otimes \rho_{S}^{(jj)} \right] (\ket{E_i} \otimes \ket{\phi}) \\
& = \bra{\phi} \rho_{S}^{(ii)} \ket{\phi} = p_i \bra{\phi} \rho_S^{(i)} \ket{\phi}.
\end{split}
\end{equation}
Since $p_i > 0$, it follows that $\bra{\phi} \rho_S^{(i)} \ket{\phi} \ge 0$ for any $\ket{\phi}$, proving positivity. Finally, recalling (\ref{trunity}), the unit trace is verified by
\begin{equation}
\mathrm{Tr} [\rho_S^{(i)}] = \mathrm{Tr} \left[ \frac{\rho_{S}^{(ii)}}{p_i} \right] = \frac{1}{p_i} \mathrm{Tr} [\rho_{S}^{(ii)}] = \frac{1}{p_i} p_i = 1.
\end{equation}

\subsection{Time evolution of $\rho_{AS}$}
\label{a2}

Assuming the structure for $H_{AS}$ given in (\ref{has}), we can derive the time evolution of $\rho_{AS}$ from (\ref{initr}) and (\ref{ras}). Since $H_{AS}$, by construction, commutes with the projectors $\ket{E_i}\bra{E_i} \otimes \mathbb{1}_S$, the evolution preserves the block-diagonal structure.

Consider the operator $ \ket{E_j} \bra{E_j} \otimes \rho_{S}^{(j)}$ and the commutator $ [ H_{AS}, \ket{E_j} \bra{E_j} \otimes \rho_{S}^{(j)}]$. Recalling the form of $H_{AS}$ in (\ref{has}), we have
\begin{equation}
    \bigl[H_{AS}, \ket{E_j} \bra{E_j} \otimes \rho_{S}^{(j)}\bigr] = 
    \ket{E_j} \bra{E_j} \otimes \bigl[ H_S + V_j, \rho_S^{(j)}\bigr].
\end{equation}
Thus, the evolution equation holds
\begin{align}
    \frac{d}{dt} \bigl[ \ket{E_j} \bra{E_j} \otimes \rho_{S}^{(j)} (t) \bigr] 
    &= - \frac{i}{\hbar} \bigl[ H_{AS}, \ket{E_j} \bra{E_j} \otimes \rho_{S}^{(j)} (t) \bigr] \nonumber \\
    &= - \frac{i}{\hbar} \ket{E_j} \bra{E_j} \otimes \bigl[ H_S + V_j, \rho_S^{(j)} (t) \bigr],
    \label{op}
\end{align}
whose solution is given by
\begin{equation}
    \ket{E_j} \bra{E_j} \otimes \rho_{S}^{(j)} (t) = 
    e^{-\frac{i H_{AS} t}{\hbar}} \bigl[ \ket{E_j}\bra{E_j} \otimes \rho_{S}^{(j)} (0) \bigr] e^{\frac{i H_{AS} t}{\hbar}}.
    \label{sol}
\end{equation}
Since $\rho_S^{(j)}$ satisfies the von Neumann equation for a suitable effective Hamiltonian $H_\textrm{eff}^{(j)}$ defined on $\mathcal{H}_S$,
\begin{equation}
    \frac{d}{dt} \rho_S^{(j) }(t) =  -\frac{i}{\hbar} \bigl[ H_\textrm{eff}^{(j)}, \rho_S^{(j)} (t)\bigr],
\end{equation}
and since $\ket{E_j} \bra{E_j}$ is time-independent, we also obtain
\begin{equation}
    \frac{d}{dt} \bigl[ \ket{E_j} \bra{E_j} \otimes \rho_S^{(j)}(t) \bigr] = -\frac{i}{\hbar} \ket{E_j} \bra{E_j} \otimes \bigl[ H_\textrm{eff}^{(j)}, \rho_S^{(j) }(t)\bigr].
    \label{op1}
\end{equation}
By comparing (\ref{op}) and (\ref{op1}), we can assume $H_\textrm{eff}^{(j)} = H_S + V_j$, and the solution of the von Neumann equation is
\begin{equation}
    \rho_S^{(j)}(t) = e^{-\frac{i(H_S + V_j)t}{\hbar} } \rho_S^{(j)}(0) e^{\frac{i(H_S + V_j)t}{\hbar} },
\end{equation}
as required by (\ref{rhoit}). 
Finally, by using (\ref{sol}), we can write the evolution of the total density matrix as given in (\ref{rast}):
\begin{align}
\rho_{AS}(t) & = e^{-\frac{i H_{AS} t}{\hbar}} \rho_{AS}(0) e^{\frac{i H_{AS} t}{\hbar}} \notag \\
& = e^{-\frac{i H_{AS} t}{\hbar}} \left[ \sum_j p_j (0) \ket{E_j}\bra{E_j} \otimes \rho_S^{(j)}(0) \right] e^{\frac{i H_{AS} t}{\hbar}} \notag \\
& = \sum_j p_j (0) \, e^{-\frac{i H_{AS} t}{\hbar}} \bigl[ \ket{E_j}\bra{E_j} \otimes \rho_S^{(j)} (0) \bigr] e^{\frac{i H_{AS} t}{\hbar}} \notag \\
&= \sum_j p_j(0) \ket{E_j} \bra{E_j} \otimes \rho_S^{(j)}(t).
\end{align}

\subsection{General short-time expansion of $\bra{E_k} \rho_{A} (t) \ket{E_l}$}
\label{a3}

Let us explicitly show the steps leading to the first order expansion of $\bra{E_k} \rho_{A} (t) \ket{E_l}$ in time, as shown in equation (\ref{elemr}), when the Hamiltonian $H_{AS}$ has the general form given in (\ref{hasgen}).

Starting from the first-order expansion in $t$ of the time evolution operator,
\begin{equation}
 e^{- i H_{AS} t / \hbar} \simeq \mathbb{1} - \frac{i H_{AS} t}{\hbar} + o(t^2), 
\end{equation}
we obtain the evolution of the density operator $\rho_{AS}$:
\begin{align}
\rho_{AS}(t) &= e^{-\frac{i H_{AS} t}{\hbar}} \rho_{AS} (0) e^{\frac{i H_{AS} t}{\hbar}} \notag \\
&\simeq (\mathbb{1} - \frac{i H_{AS} t}{\hbar}) \rho_{AS}(0) (\mathbb{1} + \frac{i H_{AS} t}{\hbar}) + o(t^2) \notag \\
&\simeq \rho_{AS}(0) + \frac{it}{\hbar} [\rho_{AS} (0) H_{AS} - H_{AS} \rho_{AS} (0)] + o(t^2). \label{eq:rho_as_expansion}
\end{align}
Now, by using 
\begin{equation} 
\rho_{AS}(0) = \ket{E_k} \bra{E_k} \otimes \rho_S^{(k)}(0)
\end{equation}
and recalling the general form $ H_{AS} = \sum_{i,j} \ket{E_i} \bra{E_j} \otimes W_{ij}$, we have
\begin{align}
\rho_{AS}(0) H_{AS}  &= \left[  \ket{E_k}\bra{E_k} \otimes \rho_S^{(k)}(0) \right] \left( \sum_{i,j} \ket{E_i} \bra{E_j} \otimes W_{ij} \right)  \notag \\
&= \sum_{i,j} (\ket{E_k}\bra{E_k}\ket{E_i}\bra{E_j}) \otimes [\rho_S^{(k)}(0) W_{ij}] \notag \\
&= \sum_{i,j}  \delta_{ki} \ket{E_k}\bra{E_j} \otimes [\rho_S^{(k)}(0) W_{ij}] \notag \\
&= \sum_{j} \ket{E_k} \bra{E_j} \otimes [\rho_S^{(k)} (0) W_{kj}]. \label{eq:rho_H_product}
\end{align}
Thus,
\begin{align}
 \rho_{AS}(0) H_{AS} (\ket{E_l} \bra{E_k} \otimes \mathbb{1}_S) &= \left[  \sum_{j} \ket{E_k} \bra{E_j} \otimes [\rho_S^{(k)} (0) W_{kj}] \right]    (\ket{E_l} \bra{E_k} \otimes \mathbb{1}_S)  \notag \\
&= \sum_{j} (\ket{E_k}\bra{E_j}\ket{E_l}\bra{E_k}) \otimes [\rho_S^{(k)}(0) W_{kj} ] \notag \\
&= \sum_{j} \delta_{jl}  \ket{E_k} \bra{E_k} \otimes [\rho_S^{(k)}(0) W_{kj}] \notag \\
&=  \ket{E_k} \bra{E_k} \otimes [\rho_S^{(k)}(0) W_{kl} ]. \label{eq:sandwich_rho_H}
\end{align}
On the other hand, since $ \rho_{AS}(0) (\ket{E_l} \bra{E_k} \otimes \mathbb{1}_S) = 0$, we have
\begin{equation}
H_{AS} \rho_{AS}(0) (\ket{E_l} \bra{E_k} \otimes \mathbb{1}_S)   =  0.
\end{equation}
Therefore, from (\ref{eq:rho_as_expansion}), the following equation holds:
\begin{equation}
\rho_{AS}(t) (\ket{E_l} \bra{E_k} \otimes \mathbb{1}_S)  \simeq  \frac{it}{\hbar} \ket{E_k} \bra{E_k} \otimes [\rho_S^{(k)}(0) W_{kl} ] + o(t^2). 
\label{sandw}
\end{equation}
Taking the trace of both sides of (\ref{sandw}) and using (\ref{expans}) (with the trace cyclicity), we finally obtain
\begin{equation}
\bra{E_k} \rho_{A} (t) \ket{E_l} \simeq \frac{it}{\hbar} \mathrm{Tr} [  W_{kl} \rho_S^{(k)}(0)] + o(t^2). 
\end{equation}
Then, for $\rho_S^{(k)}(0) = \ket{\phi} \bra{\phi}$, the expansion (\ref{elemr}) holds as desired.

\end{document}